\documentclass[a4paper,11pt]{article}
\pdfoutput=1 
\usepackage[title]{appendix}
\usepackage{jcappub} 

\usepackage{graphics}
\usepackage{xcolor,cancel}
\usepackage{caption}
\usepackage{subcaption}
\usepackage{rotating}
\usepackage{pstricks}
\usepackage{color}
\usepackage{amsfonts}
\usepackage{mathrsfs}
\usepackage{epsfig}
\usepackage{amsmath,amssymb,amsthm,graphicx,latexsym}

\definecolor{ao}{rgb}{0.0, 0.5, 0.0}
\usepackage[normalem]{ulem}
\newcommand{\stkout}[1]{\ifmmode\text{\sout{\ensuremath{#1}}}\else\sout{#1}\fi}
\definecolor{dblue}{rgb}{0,0,0.6}
\definecolor{dred}{rgb}{1,0.08,0.58}

\title{\textcolor{dblue}{\textsc{\selectfont \sffamily \bfseries Non-canonical Conformal Attractors for Single Field Inflation}}}

\author{{Tony Pinhero}}
\author{{and Supratik Pal}}

\affiliation{\textit{Physics and Applied Mathematics Unit, Indian Statistical Institute, \\203 B.T. Road, Kolkata 700 108, India}}

\emailAdd{tonypinheiro2222@gmail.com}
\emailAdd{supratik@isical.ac.in}

\abstract{We extend the idea of conformal attractors 
in inflation to non-canonical sectors by
developing a non-canonical conformally invariant 
theory from two different approaches.  In the first approach, namely, ${\cal N}=1$ supergravity,
the construction is more or less phenomenological, where the 
non-canonical kinetic 
sector is derived from a particular form of the 
K\"ahler potential respecting shift symmetry. 
In the second approach i.e., superconformal theory, we derive the form of the Lagrangian
from a superconformal action and it turns out to be exactly of the same
form
 as in the first approach.
Conformal breaking of these 
theories results in a new class of non-canonical 
models which can govern inflation with modulated shape of
the T-models. 
 We further employ this framework to explore inflationary phenomenology
 with a representative example  and show how 
 the form of the  K\"ahler potential can possibly be constrained
 in non-canonical models
using the latest   
 confidence contour 
in the $n_s-r$ plane given by recent Planck and BICEP/Keck results.}

\begin{document}
\maketitle
\flushbottom
\section{Introduction}\label{sec_intro}
The idea of spontaneous conformal/superconformal symmetry breaking in inflation \cite{kallosh_lambdaphi_4,kallosh_sup_con_gen_starob,kallosh2013universality} explains meticulously
how different class of inflationary models can 
make very similar observational predictions, 
even though their formulations are entirely 
different and their potentials are apparently 
uncorrelated. Examples include Starobinsky 
model \cite{starobinsky1980}, chaotic inflation with
$\lambda\phi^{4}$ potential and non-minimal coupling to 
gravity $\frac{\xi}{2}\phi^{2}R$ ($\xi>0$) 
\cite{salopek1989}, Higgs inflation with $\xi<0$ 
\cite{bezrukov2007} among others. With the advantage of this mechanism one can also propose new class of inflationary models \cite{kallosh2013universality} which form a universality class and in terms of observational data they all have an attractor point in the leading order approximation and these class of models are termed as conformal attractors. 

The scheme of these  conformal 
attractors  is the following: One starts with at least two 
real scalar fields. The first one is the good old inflaton field $\phi$ 
that is responsible for inflationary dynamics.
The second one(s) is(are) a conformal field(s) $\chi$, called {\it conformon}.
 These so-called conformons  
are conformally coupled to gravity and usually their kinetic 
terms are canonical, albeit with opposite sign. In addition, the potential 
terms consist of an $SO(1,1)$ symmetry breaking arbitrary 
function $F(\frac{\phi}{\chi})$ and the total action 
has a local conformal symmetry. However, as is well-known, the theory of 
inflation should not be conformally invariant. So, the way one
can make inflation  happen in the attractor framework is to
 choose a particular gauge and break the 
conformal symmetry in such a way that conformal 
field(s) get(s) decoupled from the theory. 
Thus the spontaneous symmetry breaking of 
conformal invariance results in a functional 
choice of the potential of the form 
$F(\tanh \frac{\psi}{\sqrt{6}})$ in 
Einstein frame in terms of the canonically 
normalized field $\psi$. Depending upon the 
functional choice of the potential one will 
end up with different models such as Starobinsky, 
chaotic T-models \cite{kallosh2013universality}, 
etc. 
Further, in order to realize inflation in terms of  observational data, 
one notices that
all of these models have an attractor point given by :
$1-n_{s}=\frac{2}{N_{e}}$, $r=\frac{12}{N_{e}^{2}}$ 
in the leading order approximation in $1/N_{e}$, 
where $N_{e}$ is the number of e-folding of inflation. Hence 
 the name conformal attractors. Thus, in this common
framework, idea of conformal attractors explain how 
 different, apparently uncorrelated, inflationary models end up with  identical
observational predictions. A superconformal version of the attractor scenario can further accommodate
 complex scalar fields as well but the rest of the  mechanism remains the same \cite{kallosh_superconformal1}.
This notion has further been  been
extended to multiple-field inflation scenario \cite{kallosh2013multi}, 
 non-minimal inflationary attractors \cite{kallosh_nonminimal_attra2013}, and all of
these models further been generalized to $\alpha$-attractor models \cite{kallosh2013sup_alpha_attra,kallosh2015alpha_attra}. 
The major success of these $\alpha$-attractor models are that
one can arrive at different inflationary models from a 
single Lagrangian, depending upon the different values 
of a single parameter $\alpha$ in the theory. In these class of models, kinetic term is non-canonical and it has an 
overall co-efficient $\alpha$. But in the 
potential term this parameter $\alpha$ may or may not 
appear, as it is rather a matter of choice. As a result 
the value of tensor-to-scalar ratio $r$ is proportional 
to $\alpha$ and the scalar spectral index $n_{s}$ is independent of it.

What is common among the above models is that, barring $\alpha$-attractors, all the other models deal 
with canonical fields. However, conformal breaking of non-canonical 
fields have not been well-investigated for till date.
In fact, there are very few examples  in the literature
where some non-canonical conformal invariance from some particular
superconformal theories have been studied, 
but  eventually some specific choices 
have made kinetic terms canonical  \cite{ferrara_jordanframe_sugra, ferrara_superconformalmssm, kallosh_lambdaphi_4}. 
So, proper development of conformal attractor scenario for  non-canonical fields is in need.  
On the other hand,  non-canonical models of inflation have  particularly become important
in the light of recent observational data. 
As Planck 2015 and Planck 2018 confirmed
some scale-dependence in the power spectrum at 5-$\sigma$
\cite{planck2015,planck2018_inflation}, non-canonical models, from which one can in general
generate scale-dependent power spectrum  have become more relevant than ever. 
So, this is quite timely one does a  thorough study of non-canonical conformal attractors
by investigating for proper conformal breaking of 
non-canonical models. The primary intention of the present article is 
to   
 extend the idea of conformal attractors 
to generic class of non-canonical models of 
the inflation and to see if there is any superconformal realization 
of the setup, finally leading to a demonstration of inflationary phenomenology
in the light of latest observations from Planck 2015,  2018 and BICEP/Keck  \cite{planck2015,planck2018_inflation,bicep/keck}.
In the process, we will also demonstrate how one can reproduce canonical conformal attractors
 for particular choice of the parameters
from this generic framework.

In this article, our approach is quite generic and straightforward. First,
we will start from a rather general form of 
the  non-minimal K$\ddot{a}$hler potential of $\mathcal{N}=1$ supergravity which 
is invariant under shift symmetry. We will then choose a 
particular superpotential phenomenologically, derive the non-canonical 
action therefrom, make this action conformally invariant 
by adding necessary terms into the theory. We will then employ this  K$\ddot{a}$hler potential
and superpotential to derive a generic action for inflation.
Next, we will  engage ourselves in demonstrating  how one can derive this 
apparently phenomenological action from superconformal approach.
Having establishing a proper superconformal framework for the action,
we will then develop the non-canonical conformal attractor scenario
using this action, resulting in  a generic inflaton potential.
 This generic potential is found to have 
 parametric choices  for which the  
height  of the T-model potentials 
is increasing.  The possible reason being 
higher order terms of 
non-homogeneous non-canonical kinetic term of the theory.
 Finally we employ the above framework to demonstrate briefly, with a representative example,  
inflationary phenomenology  in the 
light of latest observational data from Planck 2015, 2018, BICEP/Keck \cite{planck2015,planck2018_inflation,bicep/keck}.
We also study its 
phenomenological implications therefrom and show how one can constrain the 
K$\ddot{a}$hler potential from  observations.

\section{Basic phenomenological setup}\label{section_base_set}

Let us start with a  K\"ahler potential of the form
\begin{equation}\label{Kahler_potential}
K=-\left[\sum_{\substack{n=1}}^{l}K^{(n)}
\left(\Phi^{n}-\Phi^{*n}\right)\right]^2+SS^{*}-
\zeta (SS^*)^{2}
\end{equation}
Here $\Phi$ is the chiral superfield which plays the roll of inflation and $K^{(n)}$ are dimensionless coupling constants of their self interactions. One can notice that
this K\"ahler potential is invariant under the following shift transformation;
\begin{equation}\label{shift_symmetry}
\sum_{\substack{n=1}}^{l}K^{(n)}\Phi^{n}\rightarrow \sum_{\substack{n=1}}^{l}K^{(n)}\Phi^{n}+C^{N}
\end{equation}
This shift symmetry rather generalizes the shift symmetries proposed in \cite{kawasaki2000naturalchaotic,linear_running_kinetic_inflation_takahashi,running_kinetic_inflation_takahashi}. Due to this shift symmetry Eq.(\ref{shift_symmetry}), one can consider the term  $\sum_{\substack{n=1}}^{l}K^{(n)}\Phi^{n}$ as a composite field $\hat{\Phi}$  and the real component of this composite field $\hat{\Phi}$, will be absent in the K\"ahler potential Eq.(\ref{Kahler_potential}). Hence, the real component of $\hat{\Phi}$, i.e., $\text{Re}\left[\sum_{\substack{n=1}}^{l}K^{(n)}\Phi^{n}\right]=\text{Re}\left[\hat{\Phi}\right]$  can be identified as the inflaton scalar field. This is to avoid the usual $\eta$-problem\cite{yamaguchi2011supergravity}. However, physics is invariant under field transformation, one can also continue the same analysis  in terms of the non-canonical variable $\Phi$ of the K\"ahler potential Eq.(\ref{Kahler_potential}). So, in such a scenario, real part of $\Phi$ can be identified as inflaton, see the refs \cite{running_kinetic_inflation_takahashi,linear_running_kinetic_inflation_takahashi} for further clarifications.

The chiral multiplet $S$ plays the roll of 
an auxiliary field  and it attains a zero vev 
at the time of inflation. The term $SS^{*}$ 
gives the required potential to the $\Phi$ 
field. In absence of this, by considering the nature of 
supergravity potential and the nature of 
shift symmetry in the K$\ddot{a}$hler  
potential, the potential will not be bounded from the below\cite{kawasaki2000naturalchaotic}. 
However, if we only consider $SS^{*}$ during inflation, this field creates tachyonic instability
by acquiring a mass much smaller than the Hubble scale, resulting in the  
production of inflationary fluctuations 
and these fluctuations will be added to 
the source of isocurvature perturbations 
or to the source of non-Gaussian adiabatic 
perturbations \cite{davis2008}. This problem 
can be evaded by adding the term  $\zeta (SS^*)^{2}$  
to Eq.(\ref{Kahler_potential}), so that  the mass of the 
$S$ field will become greater than the Hubble 
scale and the corresponding fluctuations of $S$ 
will not be generated. Thus this term ensures 
the stability of inflationary trajectory near 
$S=0$. Once the stabilization is achieved, the 
field $S$ vanishes and this term becomes 
irrelevant after inflation. 
As the stabilizer field plays a crucial roll 
in the construction of supergravity inflation, 
people have investigated the nature of this 
field.   
The roll of a stabilizer field in 
supergravity inflation  and its stabilization 
issues has been discussed and explained in detail in 
\cite{kawasaki2000naturalchaotic, das2014n, Kallosh2010general_inflaton_pot_in_supergravity_stabilizerfield=sGoldstino,demozzi_supercurvaton,ferrara_jordanframe_sugra, ferrara_superconformalmssm}. 
Some recent proposals to identify this as an 
sGoldstino, a supersymmetric scalar partner 
of goldstino fermion can be found in \cite{Kallosh2010general_inflaton_pot_in_supergravity_stabilizerfield=sGoldstino}.
Alternatively, one can also 
propose a framework to 
replace stabilizer field by nilpotent superfields 
in the K$\ddot{a}$hler potential 
\cite{ferrara2014nilpotent, Kallosh2014nilpotent}. 
We will, however, consider the widely accepted stabilizer field approach. 

The K\"ahler potential defined in Eq.(\ref{Kahler_potential})
can give the following kinetic term for inflation:
\begin{multline}\label{eq_kinetic_term_in_superfields}
\frac{1}{\sqrt{-g}}L_{kin}=-2\left[\sum_{\substack{n=1}}^{l}K^{(n)}
n\Phi^{n-1}\sum_{\substack{n=1}}^{l}K^{(n)}
n\Phi^{*n-1}\right]\\
\times \partial_{\mu}\Phi
\partial^{\mu}\Phi^{*}-\left(1-4\zeta S^{*}S\right)\partial_{\mu}S\partial^{\mu}S^{*}
\end{multline}
By decomposing $\Phi$ in terms of real and imaginary components $\Phi=\frac{1}{\sqrt{2}}\left(\phi+i\chi\right)$, and by assuming along flat direction $S=\chi_{i}=0$, the Eq.(\ref{eq_kinetic_term_in_superfields}) can be written as 
\begin{equation}
\frac{1}{\sqrt{-g}}L_{kin}=-\left[\sum_{\substack{n=1}}^{l}\frac{K^{(n)}n}{\left(\sqrt{2}\right)^{n-2}}\phi^{n-1}\right]^{2}\frac{1}{2}\partial_{\mu}\phi\partial^{\mu}\phi
\end{equation}
This can be further written more simply in the form
\begin{equation}
\frac{1}{\sqrt{-g}}L_{kin}=-\left[\sum_{\substack{n=1}}^{l}l^{(n)}\phi^{n-1}\right]^{2}\frac{1}{2}\partial_{\mu}\phi\partial^{\mu}\phi
\end{equation}
with $l^{(n)}=\frac{K^{(n)}n}{\left(\sqrt{2}\right)^{n-2}}$. After squaring the series this can be written in more abstract form as
\begin{equation}
\frac{1}{\sqrt{-g}}L_{kin}=-\sum_{\substack{n=1}}^{2l-1}g^{(n)}\phi^{n-1}\frac{1}{2}\partial_{\mu}\phi\partial^{\mu}\phi
\end{equation}
In terms of new index $h$ this series can be written in more convenient form as
\begin{equation}\label{kinetic_term_from_sugra}
\frac{1}{\sqrt{-g}}L_{kin}=-\sum_{\substack{h=2}}^{2l}K^{(h)}\phi^{h-2}\frac{1}{2}\partial_{\mu}\phi\partial^{\mu}\phi
\end{equation}
Here $K^{(h)}$ is related to $g^{(n)}$ by $K^{(h)}=g^{(h-1)}$. Now for a superpotential of the form
\begin{equation}
W=S\left[\sum_{\substack{h=2}}^{2l}\left(K^{(h)}\sqrt{2}^{h}\Phi^{h}\right)^{2}\right]^{1/2}
\end{equation} 
can generate the F-term potential of the form
\begin{equation}\label{potential_from_sugra}
V=\sum_{\substack{h=2}}^{2l}\left(K^{(h)}\phi^{h}\right)^{2}
\end{equation}
Thus the total Lagrangian using Eq.(\ref{kinetic_term_from_sugra}) and Eq.(\ref{potential_from_sugra}) takes the form
\begin{equation}\label{supgravity_lagrangian}
L=\sqrt{-g}\left[\frac{R}{2}-\sum_{\substack{h=2}}^{2l}K^{(h)}\phi^{h-2}\frac{1}{2}\partial_{\mu}\phi\partial^{\mu}\phi-\sum_{\substack{h=2}}^{2l}\left(K^{(h)}\Phi^{h}\right)^{2}\right]
\end{equation}
In order to build a theory for conformal 
breaking of this non-canonical field $\phi$, we have to 
add necessary terms to the Lagrangian: first,
this field $\phi$ should be conformally coupled to $R$ and secondly, we 
have to add the conformon field $\chi$ to the Lagrangian 
in an equal footing with the $\phi$ field  in kinetic term and in coupling 
to $R$ term. By adding these terms and including 
corresponding potential term for conformon field 
in Eq.(\ref{supgravity_lagrangian}), one can write this Lagrangian in Jordan frame as
\begin{multline}\label{eq_f}
	L^{J}=\sqrt{-g}\sum_{h=2}^{2l}  \left[\frac{C^{(h)}
		\chi^{h-2}}{2}\partial_{\mu}
	\chi\partial^{\mu}\chi +\frac{C^{(h)}
		\chi^{h}}{3h^{2}}R  
	- \frac{K^{(h)}\phi^{h-2}}{2}\partial_{\mu}
	\phi\partial^{\mu}\phi \right. \\ \left. 
	-\frac{K^{(h)}\phi^{h}}{3h^{2}}R-
	\frac{4}{9h^{4}}F\left ( \frac{\phi}{\chi} \right )(K^{(h)}\phi^{h}-C^{(h)}\chi^{h})^{2} \right]
\end{multline}
where $C^{(h)}$ is the dimensionless coupling constants for the interactions of 
conformal field $\chi$. Here the term $F\left(\frac{\phi}{\chi}\right)$ have been
added to confirm quasi de Sitter evolution after the 
conformal breaking \cite{kallosh2013universality}. Thus this construction  Eq.(\ref{eq_f}) seems to be  purely phenomenological since it is well known that  the Poinca\'{r}e supergravity theories have the lack of conformon fields.
Note that for $2l=2$ this Lagrangian Eq.(\ref{eq_f})  boils down to the conformal invariant Lagrangian with canonical kinetic terms, which is used to study the construction of conformal attractors \cite{kallosh2013universality}.  Compared to this canonical conformal attractors \cite{kallosh2013universality} what we are doing is that by adding more higher order terms into the conformal attractors, we here explicitly breaking the conformal symmetry of the same. But one can notice that the Lagrangian  Eq.(\ref{eq_f}) has a conformal symmetry if one avoids the summation in $h$ index, under the following set of transformations 
\begin{equation}\label{eq_conf_transformations_in_h_index}
	g^{'}_{\mu\nu}=e^{-2\sigma(x)}g_{\mu\nu},~~~~~~~~
	\chi^{'}=e^{\frac{2}{h}\sigma(x)}\chi,~~~~~~~~
	\phi^{'}=e^{\frac{2}{h}\sigma(x)}\phi.
\end{equation}
But this observation is irrelevant in the present discussion, but what  we claim is  that one can pass through this kind of Lagrangian  Eq.(\ref{eq_f})  from a superconformal approach when the second conformal field $\eta$ is decouples from the theory, which we will show in the next section.



\section{Superconformal realization of the setup}
In the last section, we have given a phenomenological 
framework to construct a Lagrangian defined in 
Eq.(\ref{eq_f}) required for the study of conformal 
breaking of non-canonical fields. In this section we 
will demonstrate how the above Lagrangian Eq.(\ref{eq_f}) can be derived from the 
superconformal action. For this purpose, let us start our calculation by considering 
two complex scalar conformons $X_{1}^{0}$ and 
$X_{2}^{0}$, in the  K$\ddot{a}$hler embedding 
manifold ${\cal N}(X,\bar{X})$ along with scalar 
superfield $X^{1}=\Phi$ as inflaton and 
an sGoldstino $X^{2}=S$ as a stabilizer field. In 
terms of this embedding K$\ddot{a}$hler potential 
$\cal N$, general superconformal action for 
scalar-gravity part is defined  \cite{kallosh_superconformal1} as 
\begin{equation}\label{eq_s1}
\frac{1}{\sqrt{-g}}{\cal L}_{sc}^{\rm scalar-grav}=
-\frac{1}{6}{\cal N}(X,\bar{X})R-
G_{I\bar{J}}{\cal D}^{\mu}X^{I}{\cal D}_{\mu}\bar{X}^{\bar{J}}-
G^{I\bar{J}}{\cal W}_{I}\bar{\cal W}_{\bar{J}}
\end{equation}
where ${\cal W}(X)$ is the superpotential and 
$G_{I\bar{J}}$ is the K$\ddot{a}$hler metric 
which is defined as
\begin{equation}
G_{I\bar{J}}\equiv \frac{\partial^{2}{\cal N}}{\partial X^{I}\partial\bar{X}^{\bar{J}}}
\end{equation} 
and $W_{I}\equiv \frac{\partial {\cal W}}{\partial X^{I}}$ and $\bar{W}_{\bar{J}}\equiv \frac{\partial {\cal \bar{W}}}{\partial \bar{X}^{\bar{J}}}$. 
For our calculations we only need to consider 
a local conformal invariance in Eq.(\ref{eq_s1}) and need
not  bother about the other symmetries provided by 
the superconformal theory such as local special conformal 
symmetry, local $U(1)\cal R$ symmetry etc. This clearly 
indicate that, in this approach, we do not want to construct the potential 
term from the superpotential ${\cal W}(X)$. The only 
requirement on the potential term $V(X,\bar{X})$ to get 
the local conformal invariance in Eq.(\ref{eq_s1}) is 
that it should be homogeneous and second degree in both
 $X$ and $\bar{X}$ \cite{kallosh_lambdaphi_4} , which
  is stated as
\begin{equation}\label{eq_s2}
V(X,\bar{X})=\lambda^{2}\bar{\lambda}^{2}V(X,\bar{X})
\end{equation}
With this condition the superconformal action 
Eq.(\ref{eq_s1}) becomes conformal action with the scalar-gravity Lagrangian 
\begin{equation}\label{eq_s2a}
\frac{1}{\sqrt{-g}}{\cal L}_{c}^{\rm scalar-grav}=
-\frac{1}{6}{\cal N}(X,\bar{X})R-G_{I\bar{J}}{\partial}^{\mu}
X^{I}{\partial}_{\mu}\bar{X}^{\bar{J}}-V(X,\bar{X})
\end{equation}
Imposing the condition Eq.(\ref{eq_s2}) we can 
choose a potential of the form
\begin{equation}\label{eq_s3}
	V=\sum_{h=a+b}^{N}\frac{1}{9}F\left(\frac{X^{1}}{X_{1}^{0}}\right)
	\left(\frac{K^{(h)}(X^{1})^{a}(\bar{X}^{\bar{1}})^{b}}
	{E^{(h)}(X_{2}^{0})^{a-1}
		(\bar{X_{2}}^{\bar{0}})^{b-1}}-
	\frac{C^{(h)}(X_{1}^{0})^{a}(\bar{X_{1}}^{\bar{0}})^{b}}
	{E^{(h)}(X_{2}^{0})^{a-1}(\bar{X_{2}}^{\bar{0}})^{b-1}}\right )^{2}
\end{equation} 
Where $K^{(h)}$, $C^{(h)}$, and $E^{(h)}$ are the dimensionless coupling constants for the fields $X^{1}$ (inflaton superfield), $X_{1}^{0}$ (first conformon superfield)and $X_{2}^{0}$ (second conformon superfield) respectively and the values of $K^{(2)}$, $C^{(2)}$, and $E^{(2)}$ are normalized to unity. Also we consider an  embedding K$\ddot{a}$hler potential manifold
\begin{equation}\label{eq_s4}
	{\cal N}(X, \bar{X})=\left| S \right|^{2}-3\varsigma \frac{(S\bar{S})^{2}}{\left |X_{1}^{0}  \right |^{2}-
		\left | X^{1} \right |^{2}}+\sum_{h=a+b}^{N}
	\left(\frac{K^{(h)}(X^{1})^{a}
		(\bar{X}^{\bar{1}})^{b}}{E^{(h)}(X_{2}^{0})^{a-1}
		(\bar{X_{2}}^{\bar{0}})^{b-1}}-
	\frac{C^{(h)}(X_{1}^{0})^{a}(\bar{X_{1}}^{\bar{0}})^{b}}
	{E^{(h)}(X_{2}^{0})^{a-1}(\bar{X_{2}}^{\bar{0}})^{b-1}}\right )
\end{equation}

One can easily check that this embedding 
K$\ddot{a}$hler potential satisfies the 
following conditions
\begin{equation}\label{eq_s5}
{\cal N}(X,\bar{X})=X^{I}{\cal N}_{I}=
\bar{X}^{\bar{J}}{\cal N}_{\bar{J}}=X^{I}{\cal N}_{I\bar{J}}\bar{X}^{\bar{J}}
\end{equation}
where ${\cal N}_{I}\equiv\frac{\partial{\cal N}}
{\partial X^{I}}$, and
\begin{equation}\label{eq_s6}
{\cal N}_{IK\bar{L}}X^{I}={\cal N}_{\bar{I}K\bar{L}}=0, 
{\cal N}_{IK\bar{L}}\bar{X}^{L}={\cal N}_{IK}
\end{equation}
This means that ${\cal N}$ should be homogeneous and 
first degree in both X and $\bar{X}$, which implies
\begin{equation}
{\cal N}(X,\bar{X})=\lambda\bar{\lambda}({\cal N}(X,\bar{X})
\end{equation}
It has been shown in \cite{kallosh_lambdaphi_4}  
that if the K$\ddot{a}$hler potential of the 
embedding manifold ${\cal N}(X,\bar{X})$ satisfies 
the conditions Eq.(\ref{eq_s5}) and Eq.(\ref{eq_s6}) 
and the potential satisfies the condition Eq.(\ref{eq_s2}), 
then the action Eq.(\ref{eq_s2a}) has a local conformal 
invariance under the following transformations
\begin{equation}\label{eq_s7}
g^{'}_{\mu\nu}=e^{-2\sigma(x)}g_{\mu\nu},~~~~
(X^{I})^{'}=e^{\sigma(x)}X^{I},~~~~
(\bar{X}^{\bar{J}})^{'}=e^{\sigma(x)}\bar{X}^{\bar{J}}
\end{equation}

With the advantage of this conformal symmetry Eq.(\ref{eq_s7}), one can gauge away the conformal fields, which are responsible for  negative  kinetic terms, from the theory by fixing a gauge, since there are no degrees of freedom associated with these fields. The most convenient way of doing this gauge fixing is to choose a gauge which is ${\cal N}(X,\bar{X})=3$ in Eq.(\ref{eq_s1}). This helps us to recover the standard Einstein term $-\frac{R}{2}$ of supergravity in Eq.(\ref{eq_s1}). This gauge fixing can be interpreted as a migration from ${\cal N}=1$ superconformal theory to  ${\cal N}=1$ standard Poinca\'{r}e supergravity theory via spontaneous breaking of super conformal symmetry. However, for simplicity, we choose a gauge
\begin{equation}\label{eq_funda_gauge}
	{\cal N}(X,\bar{X})=-3(N-1)
\end{equation}
instead of choosing ${\cal N}(X,\bar{X})=3$. (We have explicitly checked that choosing and working with either of these two gauges will not altered the final  inflationary predictions.  If one working with the gauge ${\cal N}(X,\bar{X})=3$, then how the major equations will change is mentioned in the appendix \ref{appendix_1}. There one can see that the final inflationary predictions will not change in both the gauge choices.) This dilatational gauge Eq.(\ref{eq_funda_gauge}) can be achieved by choosing a two set of system of $N-1$ equations as follows:
\begin{equation}\label{eq_gauge_eq_1}
	E^{(h)}(X_{2}^{0})^{a-1}
	(\bar{X_{2}}^{\bar{0}})^{b-1}=\frac{h^{2}}{2}
\end{equation}
with a condition $ab\simeq\frac{h^{2}}{4}$ and
\begin{equation}\label{eq_guage_eq_2}
	C^{(h)}(X_{1}^{0})^{a}(\bar{X_{1}}^{\bar{0}})^{b}-K^{(h)}(X^{1})^{a}
	(\bar{X}^{\bar{1}})^{b}=\frac{3}{2}h^{2}.
\end{equation}
We will fix the first gauge choice Eq.(\ref{eq_gauge_eq_1}) in this section and the second gauge Eq.(\ref{eq_guage_eq_2}) in the next section. We show this gauge fixing in the theory explicitly in the following way:
In terms of ${\cal N}(X,\bar{X})$ defined 
in Eq.(\ref{eq_s4}) and in terms of potential 
defined in Eq.(\ref{eq_s3}), the action 
Eq.(\ref{eq_s2a}) reads 
\begin{multline}\label{eq_7}
\frac{1}{\sqrt{-g}}{\cal L}_{c}^{\rm scalar-grav}=
-\frac{1}{6}
\left[\sum_{h=a+b}^{N}\left(\frac{K^{(h)}(X^{1})^{a}
(\bar{X}^{\bar{1}})^{b}}{E^{(h)}(X_{2}^{0})^{a-1}
(\bar{X_{2}}^{\bar{0}})^{b-1}}-
\frac{C^{(h)}(X_{1}^{0})^{a}
(\bar{X_{1}}^{\bar{0}})^{b}}
{E^{(h)}(X_{2}^{0})^{a-1}
(\bar{X_{2}}^{\bar{0}})^{b-1}}\right )\right. \\  \left.+\left| s \right|^{2}-3\varsigma \frac{(s\bar{s})^{2}}
{\left |X_{1}^{0}  \right |^{2}
-\left | X^{1}\right|^{2}}\right]R
 -\sum_{h=a+b}^{N}\left[-G_{I\bar{J}}
 {\partial}^{\mu}X^{I}{\partial}^{\mu}\bar{X}^{\bar{J}} 
 \right. \\  \left. -\frac{1}{9}F
 \left(\frac{X^{1}}{X_{1}^{0}}\right)
 \left(\frac{K^{(h)}(X^{1})^{a}(\bar{X}^{\bar{1}})^{b}}
 {E^{(h)}(X_{2}^{0})^{a-1}(\bar{X_{2}}^{\bar{0}})^{b-1}}-
 \frac{C^{(h)}(X_{1}^{0})^{a}
 (\bar{X_{1}}^{\bar{0}})^{b}}{E^{(h)}(X_{2}^{0})^{a-1}
 (\bar{X_{2}}^{\bar{0}})^{b-1}}\right )^{2} \right ]
\end{multline}
It goes without saying that since this action has been constructed 
according to the conditions Eq.(\ref{eq_s5}),Eq.(\ref{eq_s6}) and Eq.(\ref{eq_s2}), 
it has  local conformal invariance under the above 
 transformations Eq.(\ref{eq_s7}) and the 
K$\ddot{a}$hler matrix $G_{I\bar{J}}$ takes the form
\begin{equation}
G_{I\bar{J}}\equiv\frac{\partial^{2}{\cal N}}
{\partial X^{I}\partial \bar{X}^{\bar{J}}}=
\begin{bmatrix}
 G_{0_{1}\bar{0_{1}}}& G_{0_{1}\bar{0_{2}}} 
 & G_{0_{1}\bar{1}} & G_{0_{1}\bar{2}}\\ 
G_{0_{2}\bar{0_{1}}} &  G_{0_{2}\bar{0_{2}}}
&  G_{0_{2}\bar{1}}& G_{0_{2}\bar{2}}\\ 
G_{1\bar{0_{1}}} & G_{1\bar{0_{2}}} & 
G_{1\bar{1}} &G_{1\bar{2}} \\ 
 G_{2\bar{0_{1}}}&  G_{2\bar{0_{2}}}& 
 G_{2\bar{1}} &G_{2\bar{2}} 
\end{bmatrix}
\end{equation}
We are now in a position to construct the action that we 
defined in Eq.(\ref{eq_f}). In order to do so, we will only consider 
the case in the above K$\ddot{a}$hler 
matrix $G_{I\bar{J}}$ leading to flat direction, and hence the condition is
\begin{equation}
G_{I\bar{J}}=\delta _{I\bar{J}}=
\frac{\partial^{2}{\cal N}}{\partial X^{I}
\partial \bar{X}^{\bar{J}}}
\end{equation}
Written explicitly, the components of this flat K$\ddot{a}$hler metric are as follows:
\begin{equation}
	G_{0_{1}\bar{0_{1}}}=\frac{\partial^{2}{\cal N}}
	{\partial X_{1}^{0}\partial \bar{X_{1}}^{\bar{0}}}=\sum_{h=a+b}^{N}-\frac{C^{(h)}ab(X_{1}^{0})^{a-1}
		(\bar{X_{1}}^{\bar{0}})^{b-1}}
	{E^{(h)}(X_{2}^{0})^{a-1}(\bar{X_{2}}^{\bar{0}})^{b-1}}
\end{equation}
\begin{equation}
	G_{0_{2}\bar{0_{2}}}=\frac{\partial^{2}{\cal N}}{\partial X_{2}^{0}\partial \bar{X_{2}}^{\bar{0}}}=
	\sum_{h=a+b}^{N}(1-a)(1-b)\left(\frac{K^{(h)}(X^{1})^{a}(\bar{X}^{\bar{1}})^{b}}
	{E^{(h)}(X_{2}^{0})^{a}(\bar{X_{2}}^{\bar{0}})^{b}} -\frac{C^{(h)}(X_{1}^{0})^{a}(\bar{X_{1}}^{\bar{0}})^{b}}
	{E^{(h)}(X_{2}^{0})^{a}(\bar{X_{2}}^{\bar{0}})^{b}}\right )
\end{equation}
\begin{equation}
	G_{1\bar{1}}=\frac{\partial^{2}{\cal N}}
	{\partial X^{1}\partial \bar{X}^{\bar{1}}}=
	\sum_{h=a+b}^{N}	\frac{K^{(h)}ab(X^{1})^{a-1}(\bar{X}^{\bar{1}})^{b-1}}
	{E^{(h)}(X_{2}^{0})^{a-1}(\bar{X_{2}}^{\bar{0}})^{b-1}}
\end{equation}
In terms of these components the Lagrangian Eq.(\ref{eq_s7}) 
takes the form
\begin{multline}\label{eq_s8}
	\frac{1}{\sqrt{-g}}{\cal L}_{c}^{\rm scalar-grav}=
	\sum_{h=a+b}^{N}-\frac{1}{6}
	\left[\left(\frac{K^{(h)}(X^{1})^{a}(\bar{X}^{\bar{1}})^{b}}
	{E^{(h)}(X_{2}^{0})^{a-1}(\bar{X_{2}}^{\bar{0}})^{b-1}}-
	\frac{C^{(h)}(X_{1}^{0})^{a}(\bar{X_{1}}^{\bar{0}})^{b}}{E^{(h)}(X_{2}^{0})^{a-1}
		(\bar{X_{2}}^{\bar{0}})^{b-1}}\right )R \right. \\  \left.
	+\frac{C^{(h)}ab(X_{1}^{0})^{a-1}
		(\bar{X_{1}}^{\bar{0}})^{b-1}}
	{E^{(h)}(X_{2}^{0})^{a-1}(\bar{X_{2}}^{\bar{0}})^{b-1}}
	\partial^{\mu}X_{1}^{0}\partial_{\mu}\bar{X}_{1}^{\bar{0}}-
	\frac{K^{(h)}ab(X^{1})^{a-1}(\bar{X}^{\bar{1}})^{b-1}}
	{E^{(h)}(X_{2}^{0})^{a-1}(\bar{X_{2}}^{\bar{0}})^{b-1}}
	\partial^{\mu}X^{1}\partial_{\mu}\bar{X}^{\bar{1}}\right. \\  \left.-(1-a)(1-b)\left(\frac{K^{(h)}(X^{1})^{a}(\bar{X}^{\bar{1}})^{b}}
	{E^{(h)}(X_{2}^{0})^{a}(\bar{X_{2}}^{\bar{0}})^{b}} -\frac{C^{(h)}(X_{1}^{0})^{a}(\bar{X_{1}}^{\bar{0}})^{b}}
	{E^{(h)}(X_{2}^{0})^{a}(\bar{X_{2}}^{\bar{0}})^{b}}\right )\partial^{\mu}X_{2}^{0}\partial_{\mu}
	\bar{X}_{2}^{\bar{0}}\right. \\ \left.-\frac{1}{9}F\left(\frac{X^{1}}{X_{1}^{0}}\right)
	\left(\frac{K^{(h)}(X^{1})^{a}(\bar{X}^{\bar{1}})^{b}}
	{E^{(h)}(X_{2}^{0})^{a-1}(\bar{X_{2}}^{\bar{0}})^{b-1}}-
	\frac{C^{(h)}(X_{1}^{0})^{a}(\bar{X_{1}}^{\bar{0}})^{b}}{E^{(h)}(X_{2}^{0})^{a-1}
		(\bar{X_{2}}^{\bar{0}})^{b-1}}\right )^{2}\right ]
\end{multline}
Here we have not written the terms associated with the 
stabilizer field as we know that this field attains zero 
vev during inflation. 
 If we assume the fields 
are real during inflation (corresponding partners are stabilized at zero
during inflation) 
\begin{equation}
X_{1}^{0}=\bar{X}_{1}^{\bar{0}}=\frac{\chi}{\sqrt{2}},~~~~
X_{2}^{0}=\bar{X}_{2}^{\bar{0}}=\frac{\eta}{\sqrt{2}},~~~~~ X^{1}=\bar{X}^{\bar{1}}=\frac{\phi}{\sqrt{2}}
\end{equation}
then the Lagrangian Eq.(\ref{eq_s8}) in terms of these 
real fields looks

\begin{multline}\label{eq_s9}
	{\cal L}_{c}^{\rm scalar-grav}=\sqrt{-g}\sum_{h=a+b=2}^{N}
	\left\{-\frac{R}{12}\left(\frac{K^{(h)}\phi^{h}}
	{E^{(h)}\eta^{h-2}}-\frac{C^{(h)}\chi^{h}}{E^{(h)}\eta^{h-2}} \right )+\frac{C^{(h)}ab}{2E^{(h)}}\left(\frac{\chi}{\eta} 
	\right )^{h-2}\partial^{\mu}\chi\partial_{\mu}
	\chi \right. \\  \left.-\frac{K^{(h)}ab}{2E^{(h)}}
	\left(\frac{\phi}{\eta} \right )^{h-2}\partial^{\mu}
	\phi\partial_{\mu}\phi
	-\frac{(1-a)(1-b)}{2E^{(h)}}\left[K^{(h)}\left(\frac{\phi}
	{\eta} \right )^{h}-C^{(h)}\left(\frac{\chi}{\eta} 
	\right )^{h} \right ]\partial^{\mu}\eta\partial_{\mu}
	\eta\right. \\  \left.-\frac{1}{36}F\left(\frac{\phi}{\chi}
	\right )\left(\frac{K^{(h)}\phi^{h}}{E^{(h)}\eta^{h-2}}-
	\frac{C^{(h)}\chi^{h}}{E^{(h)}\eta^{h-2}} \right )^{2} \right \}
\end{multline}

In the process, we  gauge away the second conformal 
field $\eta$ from the above theory by fixing the gauge Eq.(\ref{eq_gauge_eq_1}), and in terms of the real field $\eta$ this gauge has the form:
\begin{equation}\label{eq_gauge_eq_1_real}
	E^{(h)}\eta^{h-2}=\frac{h^{2}}{4}
\end{equation}
with an additional condition $ab\simeq\frac{h^{2}}{4}$, 
(a term $\sqrt{2}^{h}$ in the denominator of left hand side of Eq.(\ref{eq_gauge_eq_1_real}) has removed since it will cancel out with the gauge Eq.(\ref{eq_guage_eq_2}).) and  Eq.(\ref{eq_s9}) then takes the form
\begin{multline}\label{eq_s10}
L=\sqrt{-g}\sum_{h=a+b=2}^{N}  
\left[\left(\frac{C^{(h)}\chi^{h}}{3h^{2}}
-\frac{K^{(h)}\phi^{h}}{3h^{2}}\right)R+ \frac{C^{(h)}\chi^{h-2}}{2}\partial_{\mu}
\chi\partial^{\mu}\chi   
- \frac{K^{(h)}\phi^{h-2}}{2}\partial_{\mu}\phi
\partial^{\mu}\phi  \right. \\  \left.
-\frac{4}{9h^{4}}F\left ( \frac{\phi}{\chi} \right )(K^{(h)}\phi^{h}-C^{(h)}\chi^{h})^{2} \right]
\end{multline}

The resulting equation above exactly matches the erstwhile supergravity derived phenomenological model Eq.(\ref{eq_f})
proposed in the previous section under the identifications
 $N=2l$.  Thus we end
 up with a Lagrangian which contains non-canonical non-homogeneous kinetic terms for both
 conformal field $\chi$ and for inflaton field $\phi$ with an equal footing, after the decoupling of
 the second conformal field $\eta$ from the superconformal action. Now the role of this second
 conformal field $X_{2}^{0}=\frac{\eta}{2}$
  is clear, which helps us to maintain the conformal symmetry in the
 theory when one essentially wants to deal with a non-canonical non-homogeneous kinetic
 term for the conformal field $\chi$.

\section{Single field non-canonical conformal attractors}

Having convinced ourselves about the theoretical framework,
let us now employ this scenario in proposing conformal attractor framework for
non-canonical fields. For this, we start from the Lagrangian 
proposed earlier in Eq.(\ref{eq_f}). From the superconformal 
scenario, we have seen that this Lagrangian is a 
conformal broken one after the gauge fixing of 
second conformon field $\eta$. Also from the 
superconformal point of view that we have 
discussed in the previous section, this 
Lagrangian has an enhanced conformal symmetry 
when $N=2$ case, which is canonical and this 
case has studied in details in \cite{kallosh2013universality} 
and known as conformal attractors. For reader's convenience 
again we recall the Lagrangian Eq.(\ref{eq_f}) here  
\begin{equation}\label{orig_lagrang}
\begin{split}
L^{J}=\sqrt{-g}\sum_{h=2}^{N}  \left[\frac{C^{(h)}
\chi^{h-2}}{2}\partial_{\mu}
\chi\partial^{\mu}\chi 
+\frac{C^{(h)}
\chi^{h}}{3h^{2}}R  
- \frac{K^{(h)}\phi^{h-2}}{2}
\partial_{\mu}
\phi\partial^{\mu}\phi \right. \\ \left. -\frac{K^{(h)}\phi^{h}}{3h^{2}}R-
\frac{4}{9h^{4}}F
\left ( \frac{\phi}{\chi} \right )
\left(K^{(h)}\phi^{h}-C^{(h)}\chi^{h}\right)^{2} \right]
\end{split}
\end{equation}
Even though this Lagrangian has decoupled from 
second conformal field $\eta$, one can expect 
some kind of symmetry in presence of the first conformon 
field $\chi$.  As we stated earlier, we observe that this Lagrangian has a conformal symmetry if one avoids the summation in $h$ index under the transformation Eq.(\ref{eq_conf_transformations_in_h_index}). Now, we will gauge away the first conformal compensator field $\chi$ from the theory using Eq.(\ref{eq_guage_eq_2}), and in terms of the real fields this gauge choice Eq. (\ref{eq_guage_eq_2}) becomes

\begin{equation}\label{eq_guage}
C^{(h)}\chi^{h}-K^{(h)}\phi^{h}=\frac{3}{2}h^{2}
\end{equation}
Here the term $\sqrt{2}^{h}$ in the denominator of left hand side of Eq.(\ref{eq_guage})  has cancelled with the gauge choice Eq.(\ref{eq_gauge_eq_1}) in Eq.(\ref{eq_s9}).  Resolving this constraint Eq.(\ref{eq_guage}) in terms of canonically 
normalized fields $\psi_{h-1}$, one gets
\begin{equation}
\chi=\left [\frac{3h^{2}}{2C^{(h)}} 
\right ]^{\frac{1}{h}}\cosh ^{\frac{2}{h}}
\left ( \frac{\psi_{h-1}}{\sqrt{6}} \right )
\end{equation}
and
\begin{equation}\label{eq_f1}
\phi=\left [ \frac{3h^{2}}{2K^{(h)}} 
\right ]^{\frac{1}{h}}\sinh ^{\frac{2}{h}}
\left ( \frac{\psi_{h-1}}{\sqrt{6}} \right )
\end{equation}
In terms of this newly redefined canonically normalized fields $\psi_{h-1}$ the original Lagrangian Eq.(\ref{orig_lagrang}) can be expressed in Einstien frame as follows:
\begin{equation}\label{eq_g}
L^{E}=\sum_{h=2}^{N}\sqrt{-g}\left [ \frac{R}{2}-\frac{1}{2}\partial_{\mu}\psi_{h-1}
\partial^{\mu}\psi_{h-1}-F\left (\frac{\left 
[\frac{3h^{2}}{2K^{(h)}} \right ]^{\frac{1}{h}}
\sinh ^{\frac{2}{h}}\left ( \frac{\psi_{h-1}}{\sqrt{6}} 
\right )}{\left [\frac{3h^{2}}{2C^{(h)}} 
\right ]^{\frac{1}{h}}\cosh ^{\frac{2}{h}}\left ( \frac{\psi_{h-1}}{\sqrt{6}} \right )}\right )\right ]
\end{equation}
 One can readily choose $K^{(h)}=C^{(h)}$, so that the resulting 
 Lagrangian boils down to conformal attractors for $h=2$ case:
\begin{equation} \label{eq_ein}
L^{E}=\sum_{h=2}^{N}\sqrt{-g}\left [ \frac{R}{2}-\frac{1}{2}\partial_{\mu}\psi_{h-1}
\partial^{\mu}\psi_{h-1}-F\left (\tanh ^{\frac{2}{h}}
\left ( \frac{\psi_{h-1}}{\sqrt{6}} 
\right )\right )\right ]
\end{equation}
From Eq.(\ref{eq_f1}) one can easily 
see that each $\psi_{h-1}$ are not strictly independent, 
so each of these fields can be written in terms 
of a single field $\psi$ as
\begin{equation}
\psi_{h-1}=\sqrt{6}\sinh^{-1}\left[
\frac{(\sqrt{6})^{\frac{h}{2}}}{h}\sqrt{\frac{2K^{(h)}}{3}}
\sinh^{\frac{h}{2}}\left(\frac{\psi}{\sqrt{6}} \right ) \right ]
\end{equation}
It is obvious from the above equation that finally,
our framework essentially becomes an intrinsically 
single field model
in Einstein frame. One can readily check that if 
one puts this 
field back in the Lagrangian in Einstein frame 
Eq.(\ref{eq_ein}),
one readily gets back a non-canonical single 
field model.
So, we are essentially dealing with non-canonical 
conformal attractors, which is the primary target of this paper.

\begin{figure}
	\centering
	\begin{subfigure}{.45\textwidth}
		\centering
		\includegraphics[width=.8\linewidth]{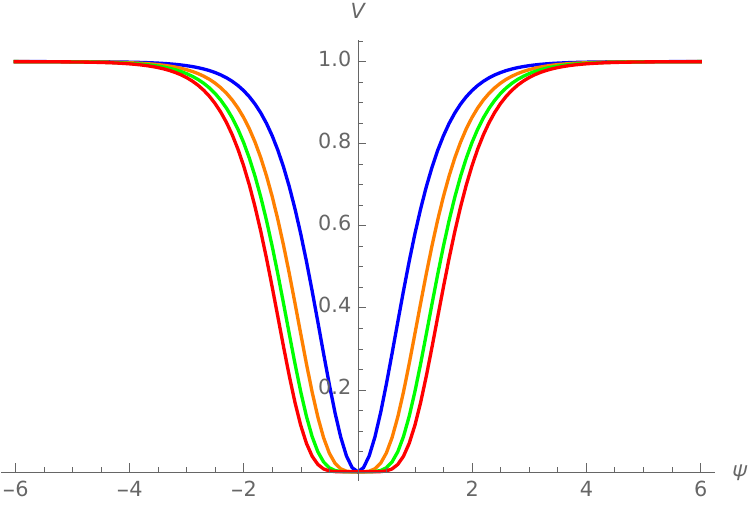}
		\caption{T-Model potentials Eq.(\ref{eq_potn}) 
			for $N=2$,  $n^{'}=1,2,3,4$ (blue, orange, green and red respectively.), and $\lambda_{n^{'}}=1$, which is proposed in \cite{kallosh2013universality}.}
		\label{fig:sub1}
	\end{subfigure}%
\hspace{1em}
	\begin{subfigure}{.45\textwidth}
	\centering
	\includegraphics[width=.8\linewidth]{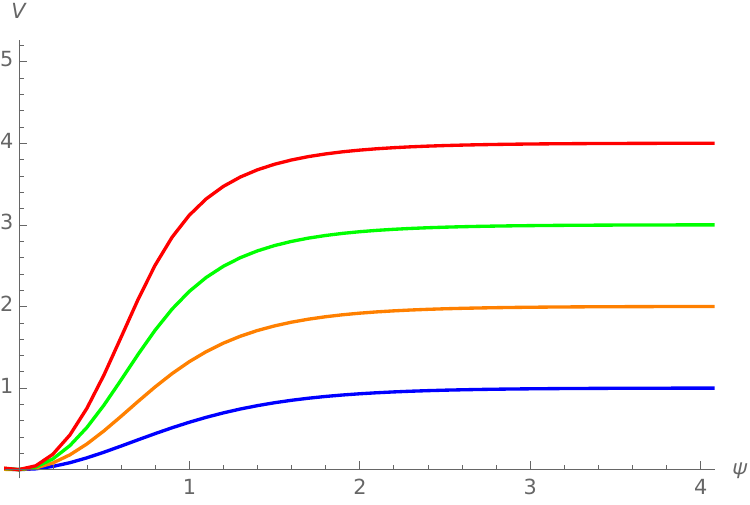}
	\caption{Behaviour of potential  Eq.(\ref{eq_potn}) in the first quadrant, for 
		different values of $N=2,3,4,5$ (blue, orange, 
		green and red  respectively.) for $\lambda_{n^{'}}=1$, $n^{'}=1$ 
		and here also we choose all $K^{(h)}=1$ for comparison. }
	\label{fig:sub3}
\end{subfigure}%

		\begin{subfigure}{.45\textwidth}
		\centering
		\includegraphics[width=.8\linewidth]{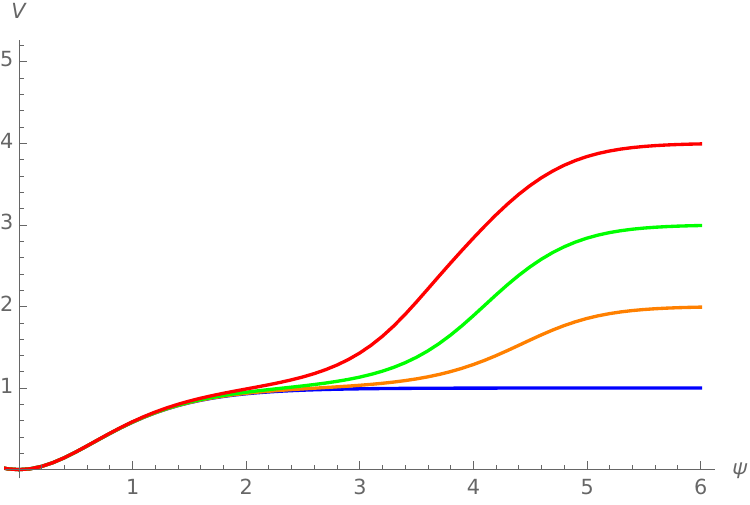}
		\caption{Effects of coupling constants $K^{(h)}$ on 
			the potential Eq.(\ref{eq_potn}) in the first quadrant, for different values of $N=2,3,4,5$ 
			(blue, orange, green and red are respectively.) 
			for $\lambda_{n^{'}}=1$, $n^{'}=1$ and for arbitrary values of $K^{(2)}=1$, $K^{(3)}=0.0000082$, $K^{(4)}=0.0000006$, $K^{(5)}=0.0000002$ }
		\label{fig:sub4}
	\end{subfigure}
\hspace{1em}
	\begin{subfigure}{.45\textwidth}
	\centering
	\includegraphics[width=.8\linewidth]{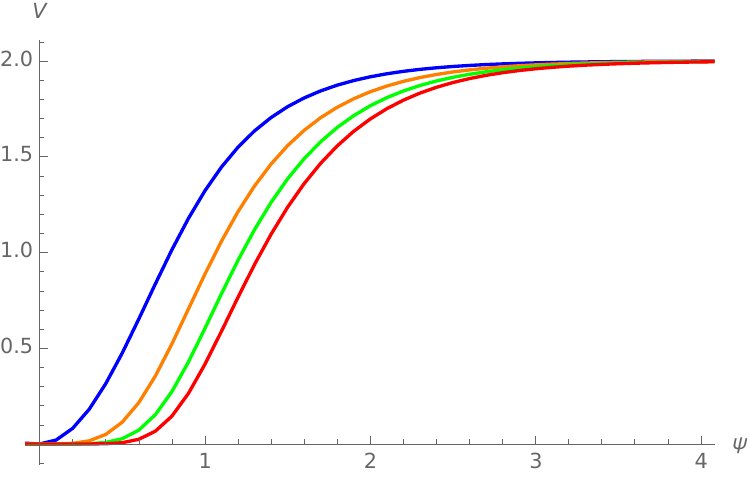}
	\caption{Behaviour of non-canonical T-model potentials Eq.(\ref{eq_potn}) in the first quadrant
		for $N=3$, $n^{'}=1,2,3,4$ 
		(blue, orange, green and red  respectively.). 
		Here we have chosen coupling constants 
		$\lambda_{n}$ and $K^{(h)}$ are unity just 
		for comparison and simplicity. }
	\label{fig:sub2}
		\end{subfigure}

	\caption{Figure depicts the potential defined in Eq.(\ref{eq_potn}) for different cases.}
	\label{fig_toy_potential_1}\label{potentials}
\end{figure}

Consequently, the final Lagrangian  takes the form
\begin{multline}\label{eq_a2}
L^{E}=\sqrt{-g}\sum_{h=2}^{N}\left \{ \frac{R}{2}-\frac{(\sqrt{6})^{h}K^{(h)}
\sinh^{h-2}(\frac{\psi}{\sqrt{6}})
\cosh^{2}(\frac{\psi}{\sqrt{6}})}{12\left[1+
\frac{2(\sqrt{6})^{h}K^{(h)}}{3h^{2}}\sinh^{h}
(\frac{\psi}{\sqrt{6}})\right]}\partial_{\mu}\psi
\partial^{\mu}\psi \right. \\  \left. -
F\left (\tanh ^{\frac{2}{h}}\left [ \sinh^{-1}\left[\frac{(\sqrt{6})^{\frac{h}{2}}}{h}
\sqrt{\frac{2K^{(h)}}{3}}\sinh^{\frac{h}{2}}
\left(\frac{\psi}{\sqrt{6}} \right ) 
\right ] \right ]\right )\right \}
\end{multline}
It is interesting to check that for $N=2$ and for $K^{(2)}=1$ in 
Eq.(\ref{eq_a2}), model becomes canonical and 
reduces to the result obtained in \cite{kallosh2013universality}
whereas for any other choice this gives rise to non-canonical conformal attractors.
Thus, one can get back canonical conformal attractors for particular choice of the parameters
from this generic framework.
This framework thus generalizes the conformal attractors as well. 
Also for $N=2$ and for $K^{(2)}=1$ in Eq.(\ref{eq_a2}),  
functional form of the potential becomes 
$F(\tanh \frac{\psi}{\sqrt{6}})$, and the simplest functional choice of the potential is
$\tanh^{2n}(\frac{\psi}{\sqrt{6}})$  and 
these are well- known T-models \cite{kallosh2013universality} and it 
has been shown in the Fig.(\ref{fig:sub1}), 
where the shape of these models symmetric 
in nature. But in our case we have to 
consider $N\geq2$ in Eq.(\ref{eq_a2}), and for the functional choice of the potential in Eq.(\ref{eq_a2}) we choose the form of the potential as follows:
\begin{equation} \label{eq_potn}
	V(\psi)=\lambda_{n}\sum_{h=2}^{N}\left 
	(\tanh ^{\frac{2}{h}}\left [ \sinh^{-1}\left[\frac{(\sqrt{6})^{\frac{h}{2}}}{h}
	\sqrt{\frac{2K^{(h)}}{3}}\sinh^{\frac{h}{2}}
	\left(\frac{\psi}{\sqrt{6}} \right ) \right ] 
	\right ]\right )^{2n^{'}}. 
\end{equation}
Again one can note that this  potential reduces to T-model for $N=2$ and for $K^{(2)}=1$ in Eq.(\ref{eq_potn}). As we are interested to
consider the effects on non-canonical fields i.e., $N\geq2$ in Eq.(\ref{eq_a2}), we consider the same in the following:
one can 
see that in the Fig.(\ref{fig:sub3}) (In this figure all coupling 
constants in the potentials taken as unity for 
comparison and simplicity which shows how this 
potential behaves depending upon the different 
values of $N$.) the height of the potential Eq.(\ref{eq_potn}) is increasing as the value of $N$ is increasing,  i.e., as the non-canonicity of the model is increasing then the height of the potential is also increasing. But if one incorporate the effects of coupling constants in the potential Eq.(\ref{eq_potn}) instead of treating all $K^{(h)}=1$, one can see that the stretching in the potentials is slightly modulated and  stretching of potential is occurring for very large values of $\psi$, which is evidenced from the Fig.(\ref{fig:sub4}). Moreover, one can observe that from the Fig.(\ref{fig:sub2}), if one fixes the non-canonicity of the model, say for $N=3$, the same broadening of the potential occurring as the values of $n^{'}$ increasing as in the case of canonical T-models. As a conclusion,
in \cite{kallosh2013universality} it has been shown 
that switching from Jordan frame to Einstein 
frame in these class of conformal models causes the exponential 
stretching of moduli space  and as a result exponential 
flattening of scalar potentials occur even if 
these potentials are very steep in the Jordan frame. Same is happening here in the case of non-canonical fields also which is evidenced from the Fig.(\ref{potentials}).

\section{Inflationary  phenomenology}\label{pheno}

Let us now employ the above framework to demonstrate briefly, with a representative example,  
inflationary phenomenology in the 
light of latest observational data from Planck 2018  and BICEP/Keck \cite{planck2018_inflation,bicep/keck}.
We will start with the particular functional choice of the potential of the form proposed in Eq.(\ref{eq_potn}).
In what follows will  mostly concentrate on the example of
 super-Planckian fields
$\psi>>1$ for which the potential Eq.(\ref{eq_potn}), and subsequently,
 the Lagrangian Eq.(\ref{eq_a2}) can be approximated as
\begin{multline}\label{eq_approx_Lagrangian}
L^{E}\simeq\sqrt{-g}\left\{\frac{R}{2}(N-1)-
\left[1+
\beta\left(1+4
e^{-\sqrt{\frac{2}{3}}\psi}\right) \right ]\frac{1}{2}\partial_{\mu}\psi\partial^{\mu}\psi-\lambda_{n}\left[(N-1)-
4n^{'}e^{-\sqrt{\frac{2}{3}}\psi} \right ] \right\}
\end{multline}
with 
\begin{equation}\label{beta_expression}
\beta=\frac{N(N+1)(2N+1)-30}{24}
\end{equation}
Now one may wonder why in this approximation suddenly 
the coupling constants disappear from the theory. 
We here remind the readers that our attempt is only to study the non-canonical effects and not to study the effects of 
coupling constants and not to constrain the values of 
these coupling constants from the observation. Even 
though  we neglect the effects of coupling constant 
in the leading approximation, we can still study 
the effects of non-canonical fields in these theories. 
After all these approximations, we can still see in the Eq.(\ref{eq_approx_Lagrangian}) the values of $N$ plays the crucial 
roll in the dynamics and this $N\geq2$ value strictly 
represents the non-canonicity in the theory.

\begin{figure}[h]
	\centering
	\includegraphics[height=10cm,width=10cm]{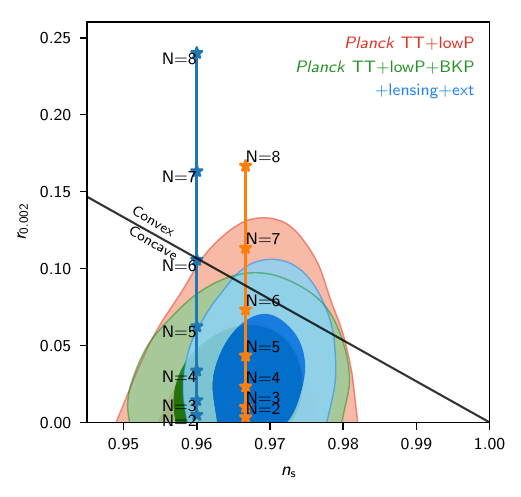}
	\caption{Comparison of the inflationary 
	observables $n_{s}$ and $r$ for the theory Eq.(\ref{eq_approx_Lagrangian}) 
 for various values of $N$ with three different confidence contours : (i) Planck TT + low P
	(red contours),  
	(ii) + BKP (green contours), and (iii) + lensing + ext (blue contours).
	The blue line corresponds to the number of e-foldings 
	$N_{e}=50$ and orange line corresponds to the  number 
	of e-folding $N_{e}=60$. The first star on the very 
	bottom of the orange line represents $N=2$ case, 
	which is the attractor point proposed in \cite{kallosh2013universality}.}\label{fig}
\end{figure}

By varying the action Eq.(\ref{eq_approx_Lagrangian}) with respect to $\psi$, one can find the equation of motion for inflaton field and at the slow-roll regime this field equation takes the form
\begin{equation}
\left[1+\beta\left(1+4e^{-\sqrt{\frac{2}{3}}\psi}\right)\right]3H\dot{\psi}=-V_{\psi}
\end{equation}
In terms of the large e-folding number $N_{e}$, this field equation can be further expressed as
\begin{equation}
\frac{d\psi}{dN_{e}}=\frac{4n^{'}}{1+\beta}\sqrt{\frac{2}{3}}\frac{1}{(N-1)}e^{-\sqrt{\frac{2}{3}}\psi}
\end{equation}
Therefore the slow-roll parameters
\begin{equation}\label{slow-roll1}
\epsilon=\frac{1}{2}\frac{1}{F(\psi)}\left(\frac{V_{\psi}}{V}\right)^{2}=\frac{3\alpha}{4N_{e}^{2}}
\end{equation}
and
\begin{equation}
\eta=\epsilon+\frac{1}{2\epsilon}\frac{d\epsilon}{dN_{e}}\simeq-\frac{1}{N_{e}}
\end{equation}
where $\alpha=1+\beta$ and $F(\psi)$ is the function in front of the kinetic term $1/2(\partial_{\mu}\psi)^{2}$ of the Lagrangian Eq.(\ref{eq_approx_Lagrangian}).
It is now straightforward to calculate the observable parameters using the above slow roll parameters.
In what follows, we only derive the two significant observable parameters
namely, the running of spectral index $n_{s}$ and the tensor-to-scalar 
ratio $r$, and confront with the confidence 
contour given by Planck 2018 and BICEP/Keck \cite{planck2018_inflation,bicep/keck}. 
Given explicitly, 
\begin{equation}\label{eq_ns}
n_{s}=1-\frac{2}{N_{e}},~~~~~~~r=\frac{12\alpha}{N_{e}^{2}},
\end{equation} 
in the leading order approximation of $1/N_{e}$. From the expression Eq.(\ref{beta_expression}), it is clear that $\alpha=1+\beta$ can only take the values $\alpha=1, 3.25, 7.25,\dots$ for the models $N=2, 3, 4,\dots$, respectively. The values of $n_{s}$ and $r$, as calculated for 
various choice of the model parameter $N$, and also for the choice of different 
number of e-foldings ${N_{e}=50}$ and ${N_{e}=60}$ ,
have been summarized in the Table (\ref{tab1}) 
and  (\ref{tab2}) respectively. The allowed regions for those parameters 
have subsequently been analyzed vis-\`{a}-vis the confidence contours  from
latest observational data in Fig.(\ref{fig}).

\begin{table}[!htb]
\begin{minipage}[c]{0.45\linewidth}
	\centering
	\begin{tabular}{|*{2}{c|}l|l|l|}
		\hline
		$\text{Sr.no}$ & $N$ & $n_{s}$ & $r$  \\
		\hline
		{1} &  {2}&  \textcolor{black}{0.96} & 
		\textcolor{black}{0.0048} \\ \hline
		{2}  &  \textcolor{black}{3} & 
		\textcolor{black}{0.96}  & 
		\textcolor{black}{0.0156}\\ \hline
		\textcolor{black}{3} &  
		\textcolor{black}{4} &  \textcolor{black}{0.96} &\textcolor{black}{0.0348} \\\hline
		\textcolor{black}{4} &  \textcolor{black}{5} & 
		\textcolor{black}{0.96} &  \textcolor{black}{0.0648}  \\ \hline
		5&6&0.96&0.108\\
		\hline
	\end{tabular}
	\captionof{table}{Values of $n_{s}$ and $r$ for 
		different $N$ values for number of e-foldings 
		$N_{e}=50$}\label{tab1}
\end{minipage}
\hfill  
\begin{minipage}[c]{0.45\linewidth}
	\centering
	\begin{tabular}{|*{2}{c|}l|l|l|}
		\hline
		$\text{Sr.no}$ & $N$ & $n_{s}$ & $r$  \\
		\hline
		{1} &  {2}&  \textcolor{black}{0.9667} &  
		\textcolor{black}{0.0033} \\ \hline
		{2}  &  \textcolor{black}{3} & 
		\textcolor{black}{0.9667}  & 
		\textcolor{black}{0.0108}\\ \hline
		\textcolor{black}{3} &  
		\textcolor{black}{4} &  
		\textcolor{black}{0.9667} &
		\textcolor{black}{0.0242} \\\hline
		\textcolor{black}{4} &  
		\textcolor{black}{5} & 
		\textcolor{black}{0.9667} &  
		\textcolor{black}{0.045}  \\ \hline
		5&6&0.9667&0.075\\
		\hline
	\end{tabular}
	\captionof{table}{Values of $n_{s}$ and $r$ for 
		different $N$ values for number of e-foldings 
		$N_{e}=60$}\label{tab2}
\end{minipage}
\end{table}

From  
Fig.(\ref{fig}) it is evident that for the number of 
e-folding $N_{e}=60$ (orange line), 
the  values of $n_{s}$  from the model fall beyond 
the 1-$\sigma$ Planck bound for $N\ge6$. This means that in the 
non-canonical sector of the theory Eq.(\ref{eq_a2}) terms 
up to
$N=5$ 
 are allowed. Inclusion of higher 
order powers of the $\psi$  in the non-canonical 
sector will lead to a tension with Planck at
1-$\sigma$, of course, for this particular choice of the potential Eq.(\ref{eq_potn}).
In other words,  when one consider 
the conformal breaking of non-canonical 
fields the terms up to $2l=N=5$ are relevant in the 
theory Eq.(\ref{eq_f}) for the corresponding 
functional choice of the potential 
$V=\left(\frac{\phi}{\chi}\right)^{2n}.$
Similarly in the case of number of e-foldings $N_{e}=50$ 
 (blue line in the Fig.(\ref{fig})), the same analysis can be done and it is evidenced that  up to $N=5$ is also allowed.  Thus, one can, in principle, 
constrain the form of non-canonical 
kinetic sector from the observation for 
the  functional choice of the 
potential defined in Eq.(\ref{eq_a2}), so as to constrain the form of K\" ahler potential defined in Eq.(\ref{eq_s4}),where the value of $N$ is allowed up to 5 in this particular case.

But recent data release from BICEP/Keck gives the bounds on the tensor to scalar ratio $r$ \cite{bicep/keck}: $r_{0.05} <0.036$ at $95 \%$ confidence. On behalf of this result one can strengthen the bounds on the non-canonicity of the model Eq.(\ref{eq_a2}) and the value of $N$ is constrained only up to $N\le4$ for both the number of e-foldings $N_{e}=50$ and $N_{e}=60$, which is evidenced from the table Table (\ref{tab1}) and Table (\ref{tab2}).

\section{Summary and Outlook}

In this article we have developed a non-canonical 
generalization of the class of conformal 
models with universal attractor behaviour 
and have also established a 
superconformal realization of the same.
We 
found that in this generalization these class 
of models  the height of the 
T-model potential is increasing due to the non-canonical 
terms in the  original conformal theory. It turns out  
 that exponential flattening of potential 
at the boundary of moduli space is occurring, when the 
fields switch from Jordan frame to Einstein 
frame though gauge fixing is violated partially 
due to the same  non-canonical terms. 

We have also engaged ourselves in finding out the phenomenological
consequences of these non-canonical conformal attractors via
a representative example for inflation. 
By confronting the values of two significant observable parameters, namely, 
the running of scalar spectral index $n_s$ and the tensor-to-scalar 
ratio $r$,  with the confidence 
contour in the $n_s-r$ plane as  given by Planck 2018 and recent BICEP/Keck,
we tried to put certain constraints on the form of the K$\ddot{a}$hler potential from observations.
It turned out that, for our particular potential under consideration, 
 in the non-canonical sector of 
the theory 
only up to $N=4$ (according to recent BICEP/Keck results \cite{bicep/keck}) terms are allowed 
when one considers number of e-foldings 
$N_{e}=60$ and the higher order terms in 
$\phi$ have to be thrown away from the kinetic 
sector due to observational constraints. 
This mechanism thus helps us put certain constraints on the 
erstwhile arbitrary K\"{a}hler potential
from observations.

 As $\alpha$-attractor models are generalized 
 version of conformal attractors, it is expected that our analysis
 for non-canonical conformal attractors should, in principle, be 
  generalized to non-canonical  $\alpha$-attractors as well.
 It is also interesting to investigate if  our single field inflation approach can be  
 extended to multi-field inflation models as well and 
 the possible consequences therefrom \cite{noncanonical multifield}.
 Further, as has been shown in a recent interesting paper \cite{alpha_dmde}, 
the $\alpha$-attractor framework can be employed to study late time phenomena like dark matter and dark energy models.
 In the same vein, it would be interesting to see the possible consequences of these
 non-canonical attractors in late time universe.
We hope to address some of these issues in near future.
 

\section*{Acknowledgments}

TP would like to thank A. Chatterjee, D. Chandra and A.Naskar 
for their fruitful suggestions and discussions. 
TP is supported by Senior Research fellowship 
(Order No. DS/16-17/0239) of the Indian Statistical 
Institute (ISI), Kolkata.
\begin{appendices}
	\section{${\cal N}(X,\bar{X})=3$ gauge}\label{appendix_1}
	If one works with the gauge ${\cal N}(X,\bar{X})=3$ instead of working with Eq.(\ref{eq_funda_gauge}), the major changes are listed below:
	
Eq.(\ref{eq_guage})	 will be modified as
	\begin{equation}\label{eq_guageA}
		C^{(h)}\chi^{h}-K^{(h)}\phi^{h}=\frac{3}{2(N-1)}h^{2}
	\end{equation}
	Resolving this constraint in terms of canonically 
	normalized fields $\psi$, one gets
	\begin{equation}
		\chi=\left [\frac{3h^{2}}{2C^{(h)}(N-1)} 
		\right ]^{\frac{1}{h}}\cosh ^{\frac{2}{h}}
		\left ( \frac{\psi_{h-1}}{\sqrt{6}} \right )
	\end{equation}
	and
	\begin{equation}\label{eq_f1A}
		\phi=\left [ \frac{3h^{2}}{2K^{(h)}(N-1)} 
		\right ]^{\frac{1}{h}}\sinh ^{\frac{2}{h}}
		\left ( \frac{\psi_{h-1}}{\sqrt{6}} \right )
	\end{equation}
Now Eq.(\ref{eq_f}) reads in Einstein frame as
\begin{equation}\label{eq_gA}
	L^{E}=\sum_{h=2}^{N}\frac{\sqrt{-g}}{(N-1)}\left [ \frac{R}{2}-\frac{1}{2}\partial_{\mu}\psi_{h-1}
	\partial^{\mu}\psi_{h-1}-\frac{1}{(N-1)}F\left (\frac{\left 
		[\frac{3h^{2}}{2K^{(h)}} \right ]^{\frac{1}{h}}
		\sinh ^{\frac{2}{h}}\left ( \frac{\psi_{h-1}}{\sqrt{6}} 
		\right )}{\left [\frac{3h^{2}}{2C^{(h)}} 
		\right ]^{\frac{1}{h}}\cosh ^{\frac{2}{h}}\left ( \frac{\psi_{h-1}}{\sqrt{6}} \right )}\right )\right ]
\end{equation}
If one chooses $K^{(h)}=C^{(h)}$,  the resulting 
Lagrangian boils down as
\begin{equation} \label{eq_einA}
	L^{E}=\sum_{h=2}^{N}\frac{\sqrt{-g}}{(N-1)}\left [ \frac{R}{2}-\frac{1}{2}\partial_{\mu}\psi_{h-1}
	\partial^{\mu}\psi_{h-1}-\frac{1}{(N-1)}F\left (\tanh ^{\frac{2}{h}}
	\left ( \frac{\psi_{h-1}}{\sqrt{6}} 
	\right )\right )\right ]
\end{equation}
 $\psi_{h-1}$ 
 fields can be written in terms 
of a single field $\psi$ as
\begin{equation}
	\psi_{h-1}=\sqrt{6}\sinh^{-1}\left[
	\frac{(\sqrt{6})^{\frac{h}{2}}}{h(N-1)^{\frac{h-2}{4}}}\sqrt{\frac{2K^{(h)}}{3}}
	\sinh^{\frac{h}{2}}\left(\frac{\psi}{\sqrt{6}} \right ) \right ]
\end{equation}
Consequently, the Lagrangian Eq.(\ref{eq_einA})  takes the form
\begin{multline}\label{eq_a2A}
	L^{E}=\sqrt{-g}\sum_{h=2}^{N}\left \{ \frac{R}{2(N-1)}-\frac{(\sqrt{6})^{h}K^{(h)}
		\sinh^{h-2}(\frac{\psi}{\sqrt{6}})
		\cosh^{2}(\frac{\psi}{\sqrt{6}})}{12(N-1)^{\frac{h}{2}}\left[1+
		\frac{2(\sqrt{6})^{h}K^{(h)}}{3h^{2}(N-1)^{\frac{h}{2}}}\sinh^{h}
		(\frac{\psi}{\sqrt{6}})\right]}\partial_{\mu}\psi
	\partial^{\mu}\psi \right. \\  \left. -\frac{1}{(N-1)^{2}}
	F\left (\tanh ^{\frac{2}{h}}\left [ \sinh^{-1}\left[\frac{(\sqrt{6})^{\frac{h}{2}}}{h(N-1)^{\frac{h-2}{4}}}
	\sqrt{\frac{2K^{(h)}}{3}}\sinh^{\frac{h}{2}}
	\left(\frac{\psi}{\sqrt{6}} \right ) 
	\right ] \right ]\right )\right \}
\end{multline}
We choose the following potential instead of  Eq.(\ref{eq_potn})
\begin{equation} \label{eq_potnA}
	V(\psi)=\frac{\lambda_{n}}{(N-1)^{2}}\sum_{h=2}^{N}\left 
	(\tanh ^{\frac{2}{h}}\left [ \sinh^{-1}\left[\frac{(\sqrt{6})^{\frac{h}{2}}}{h(N-1)^{\frac{h-2}{4}}}
	\sqrt{\frac{2K^{(h)}}{3}}\sinh^{\frac{h}{2}}
	\left(\frac{\psi}{\sqrt{6}} \right ) \right ] 
	\right ]\right )^{2n^{'}}. 
\end{equation}
For large values of $\psi>>1$ one can approximate the final Lagrangian for the Eq.(\ref{eq_a2A}) and Eq(\ref{eq_potnA}) as follows:
\begin{multline}\label{eq_approx_LagrangianA}
	L^{E}\simeq\sqrt{-g}\left\{\frac{R}{2}-
	\left[1+
	\beta\left(1+4
	e^{-\sqrt{\frac{2}{3}}\psi}\right) \right ]\frac{1}{2}\partial_{\mu}\psi\partial^{\mu}\psi-\frac{\lambda_{n}}{(N-1)}\left[1-
	\frac{4n^{'}}{(N-1)}e^{-\sqrt{\frac{2}{3}}\psi} \right ] \right\}
\end{multline}
Slow-roll equation motion for the inflaton is
\begin{equation}
	\left[1+\beta\left(1+4e^{-\sqrt{\frac{2}{3}}\psi}\right)\right]3H\dot{\psi}=-V_{\psi}
\end{equation}
In terms of the large e-folding number $N_{e}$, this field equation can be further expressed as
\begin{equation}
	\frac{d\psi}{dN_{e}}=\frac{4n^{'}}{1+\beta}\sqrt{\frac{2}{3}}\frac{1}{(N-1)}e^{-\sqrt{\frac{2}{3}}\psi}
\end{equation}
Therefore the slow-roll parameter $\epsilon$ becomes
\begin{equation}
	\epsilon=\frac{1}{2}\frac{1}{F(\psi)}\left(\frac{V_{\psi}}{V}\right)^{2}=\frac{3\alpha}{4N_{e}^{2}}
\end{equation}
This is same as that of the expression Eq.(\ref{slow-roll1}). So the rest of the inflationary analysis are same as that of the main section (\ref{pheno}) . So, it is quiet evidenced that working with the any of these two gauges will not alter the final inflationary predictions.
\end {appendices}	

\end{document}